\documentclass{article}
\usepackage{iclr2027_conference,times}
\usepackage{amsmath,amssymb,amsfonts,amsthm,bm}
\usepackage{booktabs}
\usepackage{array}
\usepackage{microtype}
\usepackage{url}
\usepackage{graphicx}
\usepackage{wrapfig}
\usepackage{capt-of}
\usepackage[table,dvipsnames]{xcolor}
\usepackage{listings}
\usepackage[
  colorlinks=true,
  linkcolor=MidnightBlue,
  citecolor=MidnightBlue,
  urlcolor=MidnightBlue
]{hyperref}

\newcolumntype{L}[1]{>{\raggedright\arraybackslash}p{#1}}

\lstdefinestyle{eqxpython}{
  language=Python,
  backgroundcolor=\color{black!3},
  basicstyle=\ttfamily\footnotesize,
  keywordstyle=\color{MidnightBlue}\bfseries,
  commentstyle=\color{ForestGreen}\itshape,
  stringstyle=\color{BrickRed},
  numberstyle=\scriptsize\color{Gray},
  numbers=left,
  numbersep=7pt,
  frame=single,
  rulecolor=\color{black!15},
  breaklines=true,
  columns=fullflexible,
  keepspaces=true,
  showstringspaces=false,
  tabsize=4
}

\title{
  Through the Looking-Glass: Efficient Parity-Complete Learning via Local $O(2)$ Frames
}

\author{Zemin Xu \\
  City University of Hong Kong, Nanjing University\\
  \texttt{xvzemin@smail.nju.edu} \\
  \And
  Wenbo Xie\thanks{Corresponding author} \\
  City University of Hong Kong \\
  \texttt{wenboxie@cityu.edu.hk} \\
}
\iclrfinalcopy

\begin{document}
\maketitle

\fancyhead{} % clear for iclr2027 preprint

\begin{abstract}
  Making all symmetry-allowed tensor representations computationally accessible remains a central challenge in equivariant atomistic learning. Clebsch--Gordan tensor products (CGTPs) are computationally demanding, whereas efficient $SO(2)$-based alternatives lack a unified treatment of natural- and unnatural-parity representations. We first introduce a generalized Wigner-$6j$ convolution that exactly recouples interactions involving additional node representations, replacing the first edge-level tensor-product intermediate with reusable node features. This reduces edge computation and storage but leaves the $\mathcal O(L^5)$ angular scaling of the sparse CGTPs unchanged. To address this remaining bottleneck, we develop a local $O(2)$ framework whose convolutions scale as $\mathcal O(L^3)$. The framework explicitly accounts for both rotations and reflections. Its linear maps, tensor products, and gated nonlinearities, combined with frame transformations, ensure global $O(3)$ equivariance for both spatial parities. Incorporating time-reversal labels extends this construction to $O(3)\times\mathbb Z_2^{\mathcal T}$. We apply the framework to magnetic TACE (mTACE), using distinct interactions for systems with and without spin--orbit coupling. On collinear CrN and noncollinear Fe benchmarks, mTACE achieves substantial reductions in atomic- and magnetic-force errors. We also provide EquivariantX, a library for global and local $O(2)$-equivariant computation.
\end{abstract}

\section{Introduction}

Modern equivariant machine learning interatomic potentials (MLIPs) have achieved high accuracy and strong transferability across chemical and structural environments, enabling simulations far beyond the reach of direct density-functional theory (DFT) calculations~\citep{NequIP,MACE,TECE}. Extending these models to additional physical degrees of freedom, however, raises a central challenge: how to retain the required symmetry information while keeping computation affordable, especially, for supporting spatial and time-reversal parities. For a maximum angular degree $L$, standard dense Clebsch-Gordan tensor-product (CGTP) evaluation has $\mathcal O(L^6)$ angular complexity. Although exploiting coefficient sparsity reduces this cost to $\mathcal O(L^5)$~\citep{oeq,cueq}, retaining both spatial parities doubles the number of irrep types relative to a natural-parity model, while tracking time-reversal parity further expands the representation space to $O(3)\times\mathbb Z_2^{\mathcal T}$. Cartesian tensor provides an alternative route, but storing a rank-$k$ Cartesian tensor in full requires $3^k$ components, also leading to exponential growth in feature storage and expensive high-rank contractions~\citep{HotPP-spin}.

Local-frame methods address this angular bottleneck by aligning features with edge directions and reducing $SO(3)$ convolutions to local $SO(2)$ operations ~\citep{eSCN,QHNetV2,TECE} . However, equivariance under rotations about a reference direction alone does not specify the action of reflections or time reversal. Spatial inversion reverses the edge direction and induces a local reflection after realignment. Features with identical local rotation laws can therefore require different reflection actions, which must be reconciled with their global spatial parities. In particular, the $m=0$ channels used for scalar nonlinearities and gating are not necessarily reflection-even or time-even. Operations permitted by $SO(2)$ symmetry can consequently mix incompatible representations or violate symmetry through their nonlinearities. The difficulty lies in consistently resolving these constraints across representation maps, linear layers, tensor products, and gates, while ensuring that the resulting local computations guarantee global $O(3)\times\mathbb Z_2^{\mathcal T}$ equivariance and retain efficient angular scaling. 

Magnetic materials provide a concrete setting in which these distinctions are essential. Their coupled atomic structure and magnetic order govern phenomena such as magnetoelasticity and spin-orbit-induced anisotropy~\citep{wu2024three,Weber2022SpinPhonon,Ding2024SpinPeierls,HotPP-spin}, with applications in information storage, energy conversion, sensing, and spin-based electronics. Atomic displacements are time-even polar vectors, whereas magnetic moments are time-odd axial vectors. For example, $\bm m_i\cdot\bm n_{ij}$ is unchanged by rotations about the edge direction $\bm n_{ij}$, but changes sign under either spatial inversion or time reversal. Although it is a scalar under local $SO(2)$ rotations, it cannot be passed through an arbitrary activation or used as an invariant gate without violating these additional symmetries. Despite recent progress in magnetic ACE, time-reversal-equivariant networks, and magnetic message passing~\citep{MagneticACE,te3nn,mMACE,HotPP-spin}, this area remains considerably less mature than conventional MLIPs.

To address these challenges, we first introduce a generalized Wigner-$6j$ convolution that exactly recouples interactions among node features, edge spherical harmonics, and additional node representations, enabling reuse of node-level intermediates across edges to reduce redundant computation. We establish a local $O(2)$ theory that preserves complete $O(3)$ interactions and extend it with time-reversal parity to realize $O(3)\times\mathbb Z_2^{\mathcal T}$ equivariance. We implement the full framework in EquivariantX, a unified library of representation maps, frame transformations, linear layers, tensor products, and gates that makes parity-complete local computation reusable across equivariant architectures. For demonstration, we adopt the TACE architecture, which shows that real-valued local weights can preserve global $O(3)$ equivariance for natural-parity features~\citep{TECE} and achieves state-of-the-art accuracy on
benchmarks such as Matbench Discovery~\citep{MatBench}. Using EquivariantX, we develop magnetic TACE (mTACE) with separate formulations for systems with and without spin-orbit coupling (SOC). mTACE achieves substantially better performance in benchmarks of magnetic systems than the reported baselines.

\section{Related Work}

\subsection{Magnetic Machine Learning Interatomic Potentials}
Early magnetic MLIPs were typically limited to collinear magnetic moments, while subsequent magnetic ACE and equivariant neural networks incorporated noncollinear magnetic moments and higher-order interactions~\citep{MagneticACE,te3nn}.  mMACE~\citep{mMACE} represents moments with spherical features and couples them to a MACE many-body expansion.  Its most expressive formulation applies one CGTP to form a geometric edge feature and a second CGTP to couple the source magnetic solid harmonics. The high cost of these successive CGTPs motivates a reduced symmetry constraint, with the practical model enforcing only $SO(3)$ equivariance. HotPP-spin~\citep{HotPP-spin} models exchange-dominated and SOC systems with a Cartesian tensor network. Coupled structural and magnetic features carry explicit spatial and time-reversal labels, which multiply under tensor couplings, while the energy readout admits only spatially even and time-even scalar channels. This construction enforces the required symmetries. Its compression of some tensor-product paths ties weights across certain irreducible components, reducing the number of independently parameterized couplings.

\subsection{Time-Reversal e3nn}
Time-reversal e3nn~\citep{te3nn} augments each $O(3)$ irrep with a $\mathbb Z_2^{\mathcal T}$ parity and propagates this label through linear maps, tensor products, gates, and other operations. It thus provides a general framework for incorporating time-reversal symmetry into CGTP-based networks, while retaining the high computational cost of CGTPs.

\subsection{Wigner-\texorpdfstring{$6j$}{6j} Convolution}
E2Former~\citep{E2Former} introduced Wigner-$6j$ recoupling as a new convolution for geometry-based energy, force, and stress prediction. It uses the binomial expansion of regular solid harmonics to rewrite $\mathcal{R}^{(\ell)}(\bm r_j-\bm r_i)$ in terms of $\mathcal{R}^{(u)}(\bm r_i)$ and $\mathcal{R}^{(\ell-u)}(\bm r_j)$, allowing angular computation to be reused across nodes. However, due to its reliance on solid harmonics, this construction is generally inequivalent to standard CGTP networks unless specifically designed to recover such equivalence. It is specialized to the positional solid-harmonic expansion and does not describe how independently supplied invariant or equivariant node inputs can directly modulate the message. 
% Moreover, although the complete expansion is algebraically translation invariant, its use of absolute positions increases finite-precision cancellation errors under global translation.

\subsection{\texorpdfstring{$SO(2)$}{SO(2)} Local Frames}
 eSCN~\citep{eSCN} showed that aligning representations with an edge reduces an $SO(3)$ convolution to $SO(2)$ operations.  QHNetV2~\citep{QHNetV2} extended this framework with $SO(2)$ tensor products and gates for Hamiltonian prediction. TECE~\citep{TECE} showed that real-valued weights preserve $O(3)$ equivariance for natural-parity features. The more challenging question here is the complete local representation needed when polar tensors, pseudotensors, and both time parities coexist.

\section{From Wigner-\texorpdfstring{$6j$}{6j} Recoupling to Local \texorpdfstring{$O(2)$}{O(2)} Frames}

A direct CGTP construction first couples a source node feature to an edge spherical harmonic, then couples this edge intermediate to an additional source-node representation, such as magnetic solid harmonics. These successive edge-level tensor products incur substantial computation and intermediate storage. Our generalized Wigner-$6j$ convolution instead couples the two node inputs once per node and reuses the result across outgoing edges. With the recoupling coefficients and path normalizations included, it exactly reproduces the two-stage CGTP without storing its first edge intermediate. Fused backends, such as OpenEquivariance~\citep{oeq} and CuEquivariance~\citep{cueq}, can thus be used to avoid explicit materialization of edge-level intermediates. Appendix~\ref{sec:wigner6j} gives the recoupling, weight construction, and backend requirements. Recoupling reduces edge-level intermediate storage but still involves a substantial number of coupling paths. Their remaining cost motivates a complementary approach: align features with a reference direction and perform the learnable update using local $O(2)$ operators. The following section establishes the rotation and reflection constraints that ensure global $O(3)$ equivariance, and then incorporates time-reversal parity.

\begin{figure}[!t]
  \centering
  \includegraphics[width=\linewidth]{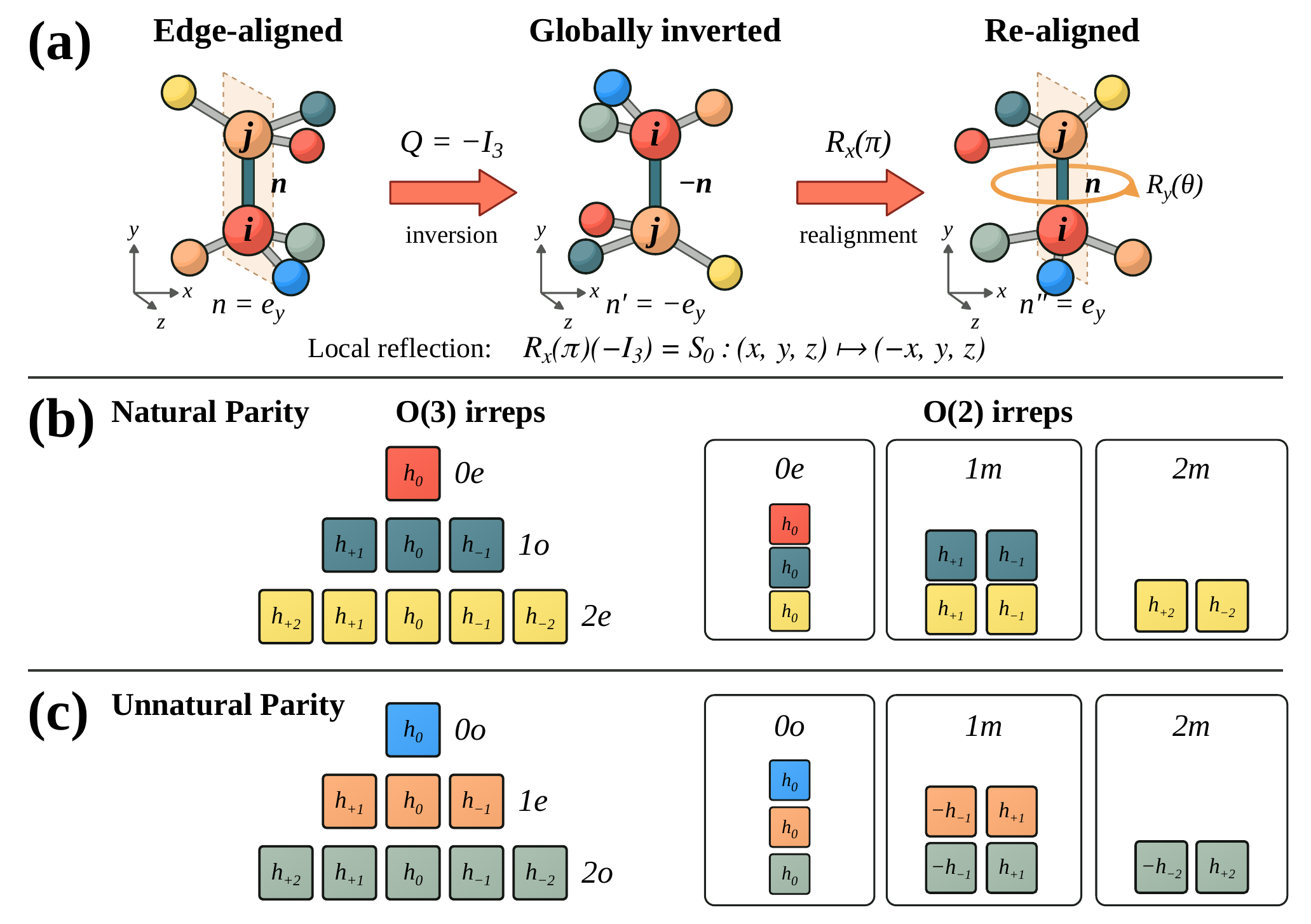}
  \caption{ Global $O(3)$ to local $O(2)$ decomposition. (a) The edge direction $\bm n$ is rotated onto the $y$ axis, leaving an $O(2)$ action in the plane perpendicular to the edge. (b) Restriction of natural-parity $O(3)$ irreps with $\ell\leq2$ into the $m=0,1,2$ local irreps. (c) The corresponding restriction of unnatural-parity irreps, where the $m=0$ irreps are reflection-odd. For $m>0$, we apply a change of basis. }
  \label{fig:o2_restriction}
\end{figure}

\section{Local \texorpdfstring{$O(2)$}{O(2)} Framework}

We label $O(3)\times\mathbb Z_2^{\mathcal T}$ irreps by $\ell pt$, where $p$ and $t$ denote spatial and time-reversal parities. Relative positions and magnetic moments transform as \texttt{1oe} and \texttt{1eo}, respectively, while the energy is an invariant \texttt{0ee} scalar. With SOC, spatial and spin rotations are coupled; without SOC, they are independent. Magnetic regular solid harmonics carry the same parities as spherical harmonics and remain smooth at zero magnetic moment. Appendix~\ref{app:notation} and ~\ref{sec:methods} summarize notation and symmetry rules. Here, we use an edge direction, given by the normalized relative position $\bm n=\bm r_{ij}/\|\bm r_{ij}\|$ (\texttt{1o}), to illustrate the construction of local $O(2)$ frames and $O(3)$-equivariant maps. The construction is analogous for axial reference directions (\texttt{1e}), but the stabilizer subgroup is $SO(2)\times\mathbb Z_2$.

\subsection{Local frames and stabilizer}

Choose a proper rotation $R_{\bm n}\in SO(3)$ that aligns $\bm n$ with $\bm e_y=(0,1,0)^{\mathsf T}$:
\begin{equation}
  R_{\bm n}\bm n=\bm e_y.
  \label{eq:appendix-frame-condition}
\end{equation}
The stabilizer of the aligned direction is
\begin{equation}
  H=\{g\in O(3):g\bm e_y=\bm e_y\}\simeq O(2) \simeq SO(2)\rtimes\mathbb Z_2.
  \label{eq:appendix-isotropy-subgroup}
\end{equation}
Its elements act orthogonally on the perpendicular $xz$ plane. Those with determinant $+1$ are rotations about the $y$ axis; those with determinant $-1$ are reflections through planes containing the $y$ axis.
A global transformation $Q\in O(3)$ acts on both the features and $\bm n$. The corresponding transformation in the aligned frame is
\begin{equation}
  g(Q,\bm n)=R_{Q\bm n}Q R_{\bm n}^{-1},
  \qquad
  g(Q,\bm n)\bm e_y=R_{Q\bm n}Q\bm n=\bm e_y,
  \qquad
  \det g(Q,\bm n)=\det Q.
  \label{eq:appendix-frame-cocycle}
\end{equation}
By definition, $g(Q,\bm n)R_{\bm n}=R_{Q\bm n}Q$: first applying the global transformation $Q$ and then aligning is equivalent to first aligning and then applying the corresponding local transformation $g(Q,\bm n)$. The condition $g(Q,\bm n)\bm e_y=\bm e_y$ shows that $g(Q,\bm n)\in H$: it leaves the aligned direction unchanged and acts as an $O(2)$ transformation on the perpendicular $xz$ plane. Since both alignment rotations are proper, $\det g(Q,\bm n)=\det Q$. Thus proper and improper global transformations induce local rotations and reflections, respectively. Local reflection equivariance is therefore required for global $O(3)$ equivariance. The group $H$ is generated by rotations $R_y(\theta)$ and one reflection $S_0$. Every reflection in $H$ has the form $R_y(\theta)S_0$. We choose
\begin{equation}
  S_0=R_x(\pi)(-I_3)=
  \begin{pmatrix}
    -1&0&0\\
    0&1&0\\
    0&0&1
  \end{pmatrix}.
\end{equation}
This reflects positions through the $yz$ plane, or equivalently across the $z$ axis in the $xz$ plane.  Although inversion $-I_3$ is not in $H$, it induces a local reflection after realignment. For example, the valid alignments $R_{\bm e_y}=I_3$ and $R_{-\bm e_y}=R_x(\pi)$ give $g(-I_3,\bm e_y)=S_0$.

\subsection{Restriction and local basis}

\paragraph{Rotations.}
Let $\rho_{\ell,p}$ denote the real $O(3)$ irrep of degree $\ell$ and spatial parity $p$. We use a real spherical harmonic basis with angular factors $\cos m\phi$ and $\sin m\phi$ for $h_{+m}$ and $h_{-m}$, respectively, where $\phi$ is measured from $z$ toward $x$. Rotations about $y$ leave $h_0$ invariant and act on each two-dimensional subspace with $m=1,\ldots,\ell$ as
\begin{equation}
  \begin{pmatrix}h_{+m}\\h_{-m}\end{pmatrix}
  \mapsto R_m(\theta)\begin{pmatrix}h_{+m}\\h_{-m}\end{pmatrix},
  \qquad
  R_m(\theta)=
  \begin{pmatrix}
    \cos m\theta&-\sin m\theta\\
    \sin m\theta&\cos m\theta
  \end{pmatrix}.
\end{equation}
These proper rotations are independent of $p$.

\paragraph{Reflection and change of basis.}
The rotation matrices above are independent of spatial parity, but the reflection action is not. In the basis defined above, $S_0=(-I_3)R_x(\pi)$ gives
\begin{equation}
  \rho_{\ell,p}(S_0)
  =pD^{(\ell)}(R_x(\pi))
  =p(-1)^\ell\left(1\oplus\bigoplus_{m=1}^{\ell}F\right),
  \qquad
  F=\begin{pmatrix}1&0\\0&-1\end{pmatrix}.
\end{equation}
The factor $p$ comes from spatial inversion, while $(-1)^\ell$ follows from the Wigner-$D$ matrix in this basis. At $m=0$, the reflection character $q=p(-1)^\ell$ distinguishes the one-dimensional irreps $(0,+1)=\texttt{0e}$ and $(0,-1)=\texttt{0o}$. Their opposite signs cannot be related by a one-dimensional change of basis. For $m>0$, natural- and unnatural-parity $O(3)$ irreps therefore give the same rotation matrix $R_m(\theta)$ but opposite reflection matrices $F$ and $-F$. This difference is inherited from the parent $O(3)$ irreps, so their features cannot be mixed directly as channels of the same local irrep. We leave the natural-parity features unchanged and transform those with $p(-1)^\ell=-1$ by
\begin{equation}
  \begin{gathered}
    J=\begin{pmatrix}0&-1\\1&0\end{pmatrix},
    \qquad
    \begin{pmatrix}h'_{+m}\\h'_{-m}\end{pmatrix}
    =J\begin{pmatrix}h_{+m}\\h_{-m}\end{pmatrix}
    =\begin{pmatrix}-h_{-m}\\h_{+m}\end{pmatrix},\\
    JR_m(\theta)J^{-1}=R_m(\theta),
    \qquad J(-F)J^{-1}=F.
  \end{gathered}
\end{equation}
After this change of basis, features with the same $m>0$ use identical rotation and reflection matrices and can be mixed across channels. We denote this two-dimensional real irrep by $(m,0)$, where $0$ indicates the absence of an independent reflection-parity label.

\paragraph{Restriction rule.}
Under $H\simeq O(2)$, an $O(3)$ irrep decomposes as
\begin{equation}
  \left.(\ell,p)_{O(3)}\right\downarrow_{O(2)}
  =
  \underbrace{\bigl(0,p(-1)^\ell\bigr)}_{\text{1D},\,m=0}
  \oplus
  \bigoplus_{m=1}^{\ell}\underbrace{(m,0)}_{\text{2D}}.
  \label{eq:restriction}
\end{equation}
Figure~\ref{fig:o2_restriction} illustrates the restriction for O(3) natural- and unnatural-parity irreps.

\subsection{Global equivariance}
\label{sec:o2-theory}

Let $\rho_{\rm in}$ and $\rho_{\rm out}$ be direct sums of $O(3)$ irreps for the input and output features, respectively. Let $B_{\rm in}$ and $B_{\rm out}$ be the fixed orthogonal transformations into the local $O(2)$ basis: they apply $J$ to every positive-order subspace with $p(-1)^\ell=-1$, and leave the remaining subspaces unchanged. In this basis, the restricted representations are
\begin{equation}
  \sigma_{\rm in}(g)=B_{\rm in}\rho_{\rm in}(g)B_{\rm in}^{-1},
  \qquad
  \sigma_{\rm out}(g)=B_{\rm out}\rho_{\rm out}(g)B_{\rm out}^{-1},
  \qquad g\in H.
\end{equation}
Let $\Phi$ be an $O(2)$-equivariant operator in this basis:
\begin{equation}
  \Phi\!\left(\sigma_{\rm in}(g)h\right)
  =\sigma_{\rm out}(g)\Phi(h),
  \qquad g\in H.
  \label{eq:appendix-local-equivariance}
\end{equation}
The resulting learnable map
\begin{equation}
  \mathcal F_{\bm n}(h)
  =\rho_{\rm out}(R_{\bm n}^{-1})B_{\rm out}^{-1}
  \Phi\!\left(B_{\rm in}\rho_{\rm in}(R_{\bm n})h\right)
  \label{eq:appendix-lifted-edge-map}
\end{equation}
is then globally equivariant:
\begin{equation}
  \mathcal F_{Q\bm n}\!\left(\rho_{\rm in}(Q)h\right)
  =\rho_{\rm out}(Q)\mathcal F_{\bm n}(h),
  \qquad Q\in O(3).
  \label{eq:appendix-global-equivariance-result}
\end{equation}

\paragraph{Proof.}
With $g=g(Q,\bm n)\in H$ from Eq.~\eqref{eq:appendix-frame-cocycle}, the frame transformations satisfy
\begin{equation}
  \begin{aligned}
    B_{\rm in}\rho_{\rm in}(R_{Q\bm n})\rho_{\rm in}(Q)
    &=B_{\rm in}\rho_{\rm in}(g)\rho_{\rm in}(R_{\bm n})
    =\sigma_{\rm in}(g)B_{\rm in}\rho_{\rm in}(R_{\bm n}),\\
    \rho_{\rm out}(R_{Q\bm n}^{-1})B_{\rm out}^{-1}\sigma_{\rm out}(g)
    &=\rho_{\rm out}(R_{Q\bm n}^{-1})\rho_{\rm out}(g)B_{\rm out}^{-1}
    =\rho_{\rm out}(Q)\rho_{\rm out}(R_{\bm n}^{-1})B_{\rm out}^{-1}.
  \end{aligned}
  \label{eq:appendix-wigner-intertwining}
\end{equation}
Applying these identities together with local equivariance gives
\begin{equation}
  \begin{aligned}
  \mathcal F_{Q\bm n}\!\left(\rho_{\rm in}(Q)h\right)
  &=\rho_{\rm out}(R_{Q\bm n}^{-1})B_{\rm out}^{-1}
  \Phi\!\left(\sigma_{\rm in}(g)B_{\rm in}\rho_{\rm in}(R_{\bm n})h\right)\\
  &=\rho_{\rm out}(R_{Q\bm n}^{-1})B_{\rm out}^{-1}\sigma_{\rm out}(g)
  \Phi\!\left(B_{\rm in}\rho_{\rm in}(R_{\bm n})h\right)\\
  &=\rho_{\rm out}(Q)\mathcal F_{\bm n}(h).
  \end{aligned}
\end{equation}
\paragraph{Independence of the alignment choice.}
Any other proper alignment of the same direction has the form $R'_{\bm n}=R_y(\theta)R_{\bm n}$. Substituting it into Eq.~\eqref{eq:appendix-lifted-edge-map} inserts $\sigma_{\rm in}(R_y(\theta))$ before $\Phi$ and $\sigma_{\rm out}(R_y(\theta))^{-1}$ after it. These transformations cancel by Eq.~\eqref{eq:appendix-local-equivariance}, giving $\mathcal F'_{\bm n}(h)=\mathcal F_{\bm n}(h)$. Thus the output is independent of the arbitrary rotation of the frame about the aligned direction, including direction-dependent choices of $\theta$.

\subsection{EquivariantX for \texorpdfstring{$O(2)$}{O(2)}-Equivariant Operators}

\begingroup
\setlength{\intextsep}{\baselineskip}
\setlength{\abovecaptionskip}{\baselineskip}
\setlength{\belowcaptionskip}{\baselineskip}
\begin{wrapfigure}{R}{0.5\columnwidth}
  \centering
  \begin{minipage}[t]{\linewidth}
    \centering
    \includegraphics[width=\linewidth]{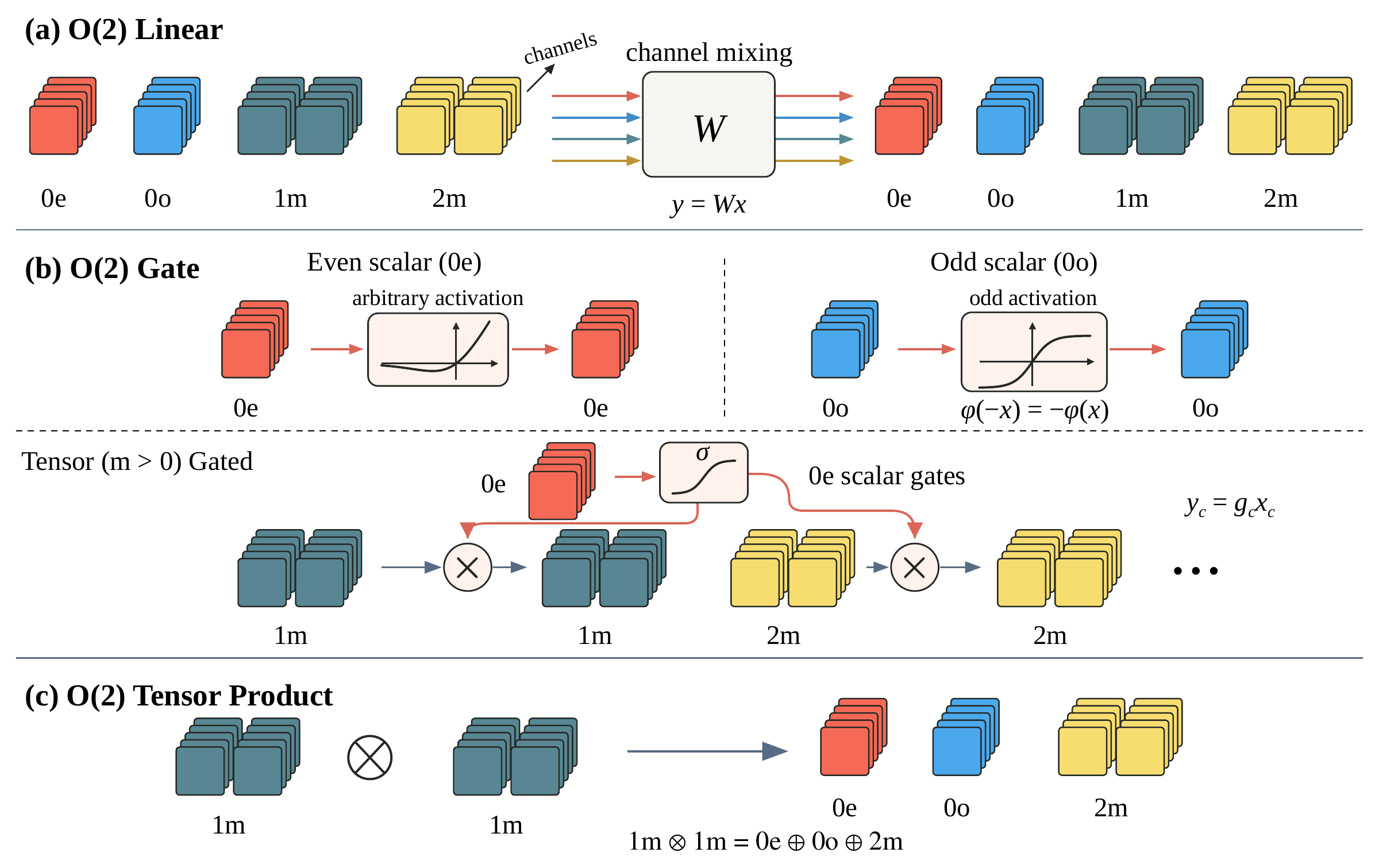}
    \caption{Core $O(2)$-equivariant operations in EquivariantX: (a) Linear transformations. (b) Gated nonlinearities. (c) Tensor products.}
    \label{fig:o2}
\par
  \centering
    \captionof{table}{Spatial tensor-product rules for real $O(2)$ irreps.}
    \label{tab:o2-tensor-product}
    \resizebox{\linewidth}{!}{
      \begin{tabular}{ll}
        \toprule
        Input irreps                                          & Output irreps \\
        \midrule
        \texttt{0e}$\otimes$\texttt{0e}                       & \texttt{0e} \\
        \texttt{0e}$\otimes$\texttt{0o}                       & \texttt{0o} \\
        \texttt{0o}$\otimes$\texttt{0o}                       & \texttt{0e} \\
        \texttt{0e}$\otimes$\texttt{$m$m}                     & \texttt{$m$m} \\
        \texttt{0o}$\otimes$\texttt{$m$m}                     & \texttt{$m$m} \\
        \texttt{$m_1$m}$\otimes$\texttt{$m_2$m}, $m_1\ne m_2$ & \texttt{$|m_1-m_2|$m}$\oplus$\texttt{$(m_1+m_2)$m} \\
        \texttt{$m$m}$\otimes$\texttt{$m$m}                   & \texttt{0e}$\oplus$\texttt{0o}$\oplus$\texttt{$2m$m} \\
        \bottomrule
      \end{tabular}
    }

  \end{minipage}
\end{wrapfigure}

EquivariantX adopts a flattened \texttt{ir\_mul} layout and provides the frame transforms above together with native $O(2)$ linear maps, gates, and tensor products. These operators are equivariant under the rotation and reflection representations of their declared irreps and can also be used independently of a local frame. Figure~\ref{fig:o2_restriction} and Figure~\ref{fig:o2} summarize them; Appendix~\ref{sec:eqx-example} gives an executable example.
\textbf{O(2) Local Frame.} We establish the transformation between global $O(3)$ features and local $O(2)$ features in a form compatible with the \texttt{e3nn}~\citep{e3nn}.
\textbf{O(2) Linear.} Linear mixes identical irreps using real-valued weights. A bias is allowed only for \texttt{0e}.
\textbf{O(2) Gate.} For label-preserving activations, \texttt{0e} admits arbitrary nonlinearities, while \texttt{0o} requires odd ones. Positive-order irreps are multiplied by \texttt{0e} gates. Since local scalars may arise from $\ell>0$ irreps, the nonlinearities are not restricted to $\ell=0$ features.
\mbox{\textbf{O(2) Tensor Product.}} Tensor products decompose according to Table~\ref{tab:o2-tensor-product}. In particular, coupling a \texttt{0o} scalar $s$ to a positive-order feature $h_m=(h_{+m},h_{-m})^{\mathsf T}$ uses the map $sJh_m$, rather than direct scalar multiplication, to retain the reflection convention $F$.

\subsection{Time-reversal symmetry}

A time-even reference direction leaves the frame unchanged under time reversal. The time label $t$ is therefore preserved after restriction:
\begin{equation}
  \left.(\ell,p,t)_{O(3)\times\mathbb Z_2^{\mathcal T}}\right\downarrow_{O(2)\times\mathbb Z_2^{\mathcal T}}
  =\bigl(0,p(-1)^\ell,t\bigr)\oplus\bigoplus_{m=1}^{\ell}(m,0,t).
\end{equation}
Requiring $\Phi$ to be $O(2)\times\mathbb Z_2^{\mathcal T}$-equivariant gives global $O(3)\times\mathbb Z_2^{\mathcal T}$ equivariance by the same proof. Linear maps preserve $t$, and tensor products multiply it. Label-preserving scalar activations must be odd if $q=-1$ or $t=-1$; invariant gates and biases carry $(m,q,t)=(0,+1,+1)$. Time-reversal e3nn~\citep{te3nn} and EquivariantX supply the global and local operators, respectively.

\par
\endgroup

\section{Magnetic TACE}
\label{sec:model}

\begin{figure}[!b]
  \centering
  \includegraphics[width=\linewidth]{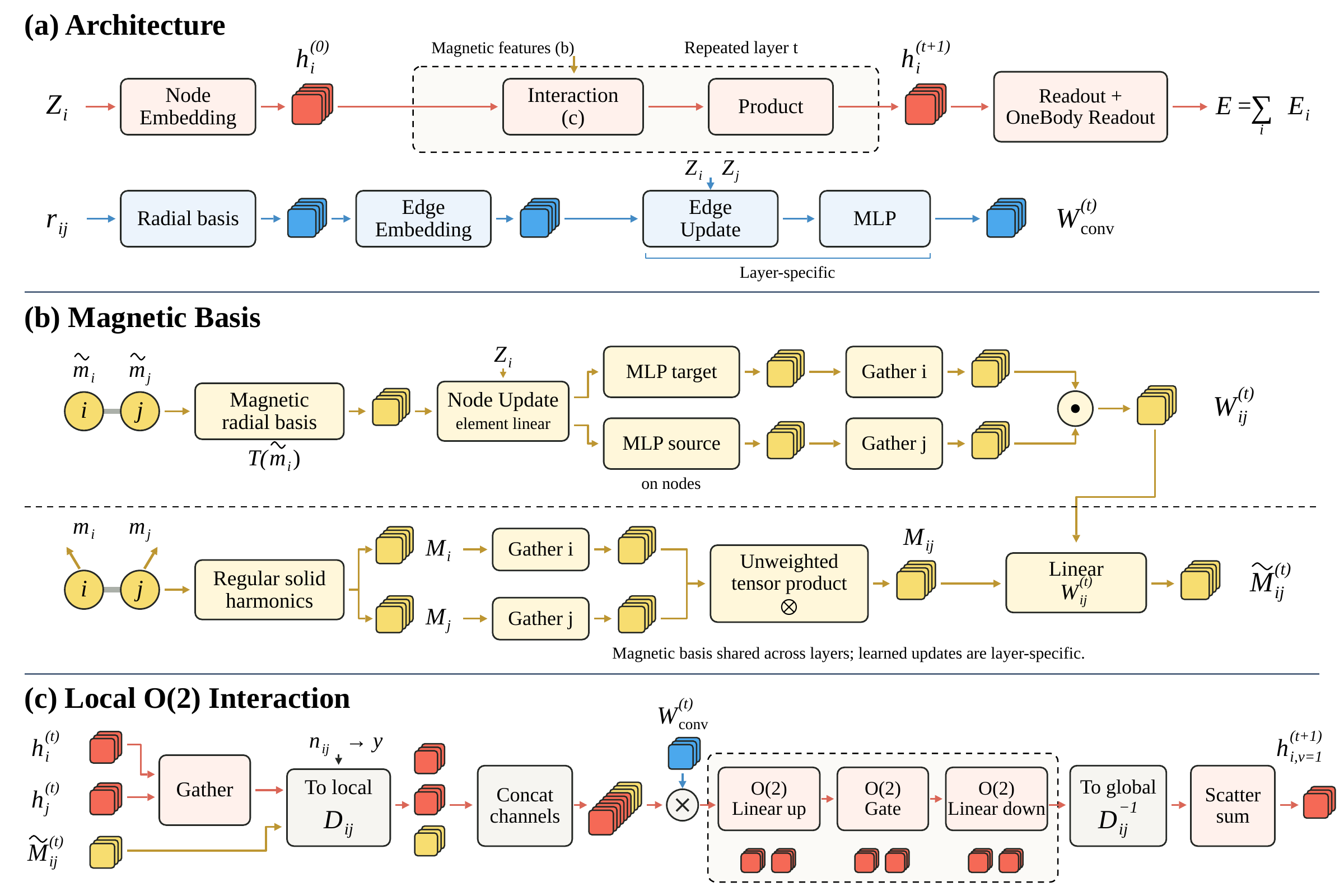}
  \caption{mTACE architecture. (a) Model overview with embedding, update, interaction, product, and readout blocks. (b) Construction of magnetic edge attributes and layer-specific radial weights. (c) A minimal magnetic local $O(2)$ convolution, where $\nu=1$ denotes correlation order one before the product-basis expansion.}
  \label{fig:arch}
\end{figure}

mTACE augments every spatial edge with magnetic information.  Unlike positions, magnetic moments are unchanged by a translation.  Therefore, we evaluate the angular basis on nodes and use regular solid harmonics to ensure smoothness at $\bm m_i=\bm 0$.  The magnetic cutoff $m_{\mathrm{cut},Z}$ and basis normalization are described in Appendix~\ref{sec:magnetic-normalization}.

\paragraph{SOC and non-SOC Magnetic Local $O(2)$ Convolution.} The SOC model is capable of capturing the full range of interactions. However, generating SOC training data is computationally expensive. Therefore, our model is designed to provide a fully symmetry-preserving SOC formulation together with a compatible non-SOC counterpart. This unified framework makes it possible to pretrain foundation models on large-scale, relatively inexpensive non-SOC datasets and subsequently fine-tune them on smaller SOC datasets. Both variants use the magnetic node attributes $\mathcal M_i=\bigoplus_{\ell=0}^{L_{\rm mag}}\mathcal R^{(\ell)}(\bm m_i)$ and differ in the magnetic output degrees retained on each edge:
\begin{equation}
  \mathcal M_{ij}^{\rm SOC}=
  \bigoplus_{\substack{0\le \ell_i,\ell_j\le L_{\rm mag}\\|\ell_i-\ell_j|\le L\le\min(\ell_i+\ell_j,L_{\rm mag})}}
  \left[\mathcal M_i^{(\ell_i)}\otimes
  \mathcal M_j^{(\ell_j)}\right]_{L},\;
  \mathcal M_{ij}^{\mathrm{non\text{-}SOC}}=
  \bigoplus_{\ell=0}^{L_{\rm mag}}
  \left[\mathcal M_i^{(\ell)}\otimes
  \mathcal M_j^{(\ell)}\right]_{L=0}.
  \label{eq:magnetic-edge-attributes}
\end{equation}
Here $\ell_i,\ell_j$ are node angular degrees and $L$ is the magnetic edge output degree, all truncated at $L_{\rm mag}$. Both products are unweighted, and coupling paths remain separate. The non-SOC architecture retains only $L=0$, which forces $\ell_i=\ell_j$ and gives $(L_{\rm mag}+1)$ copies of \texttt{0ee}.  These channels preserve exchange correlations, including $\bm m_i\!\cdot\!\bm m_j$, while remaining invariant under spin rotations.  The SOC architecture additionally retains non-scalar channels that can couple spin orientation to the lattice. Let $\widetilde m_i$ denote the transformed magnetic magnitude.  In layer $t$, the magnetic node features $T(\widetilde m_i)$, where $T$ denotes the Chebyshev polynomials of the first kind, are mapped by an element-dependent linear embedding and then passed through independent source and target MLPs on nodes. Their outputs are gathered onto each edge and multiplied element-wise ($\odot$) to generate the weights for subsequent linear layers that project all irreps to the same number of channels.
\begin{equation}
  W_{ij}^{(t)}=\operatorname{MLP}_{\rm tgt}^{(t)}\!\left(
  W_{Z_i}^{(t)}T(\widetilde m_i)\right)
  \odot\operatorname{MLP}_{\rm src}^{(t)}\!\left(
  W_{Z_j}^{(t)}T(\widetilde m_j)\right),
  \quad
  \widetilde{\mathcal M}_{ij}^{(t)}
  =\operatorname{Linear}\!\left(
  \mathcal M_{ij};
  W_{ij}^{(t)}
  \right).
  \label{eq:magnetic-edge-update}
\end{equation}
For each edge $(i,j)$, we gather the node features $h_i^{(t)}$ and $h_j^{(t)}$ and combine them with the magnetic edge features $\widetilde{\mathcal M}_{ij}^{(t)}$ from Eq.~\eqref{eq:magnetic-edge-update}. Writing $D_{ij}$ for the representation-specific Wigner-$D$ transform into edge-aligned frames, including the fixed local basis change above, the local input is
\begin{equation}
  h_{ij}^{(t)}=D_{ij}h_i^{(t)}\oplus D_{ij}h_j^{(t)}
  \oplus D_{ij}\widetilde{\mathcal M}_{ij}^{(t)}.
  \label{eq:local-magnetic-inputs}
\end{equation}
The $O(2)$ convolution used in all benchmark models is
\begin{equation}
  h_{ij}^{(t)}
  \xrightarrow{\ \bm W_{\text{conv},ij}^{(t)}\ }
  O(2)\text{ Linear}_{\rm up}
  \longrightarrow O(2)\text{ Gate}
  \longrightarrow O(2)\text{ Linear}_{\rm down}
  \longrightarrow\operatorname{RRA},
  \label{eq:o2-magnetic-convolution}
\end{equation}
where $\bm W_{\text{conv},ij}^{(t)}$ are generated from spatial radial features and the source--target element embedding. All benchmark models use Radial Rotary Complex Attention (RRA)~\citep{TECE} after the final Linear to incorporate radial information. The message is rotated back to the global frame and scattered to the target node.  The subsequent TACE product basis performs the explicit many-body expansion, and the energy readout retains only the invariant \texttt{0ee} scalar. See Fig.~\ref{fig:arch} for the complete architecture of $O(2)$-mTACE. The Wigner-$6j$ interaction is described separately in Appendix~\ref{sec:wigner6j} and is not used in these benchmarks.

\begin{table}[t]
  \centering
  \caption{RMSEs for the magnetic benchmarks.  Energy, atomic-force, stress, and magnetic-force errors are denoted by $E$, $F$, $S$, and $F^{\mathrm{mag}}$ and reported in meV/atom, meV/\AA, meV/\AA$^3$, and meV/$\mu_{\mathrm B}$, respectively. A dash means that no result was reported. The best results are highlighted in \textbf{bold}.}
  \label{tab:magnetic-benchmarks}
  \small
  \begin{minipage}[t]{0.49\linewidth}
    \centering
    \colorbox{green!10}{\strut\textbf{Collinear CrN}}\\[3pt]
    \scriptsize
    \setlength{\tabcolsep}{1pt}
    \begin{tabular*}{\linewidth}{@{\extracolsep{\fill}}llrrrr}
      \toprule
      Model      & Type                   & $E$  & $F$   & $S$  & $F^{\mathrm{mag}}$ \\
      \midrule
      mMTP       & non-SOC                & 1.7  & 108   & 2.50 & 64 \\
      mMACE      & SOC\textsuperscript{a} & 1.21 & 24    & 1.56 & 19 \\
      mTACE & non-SOC                & 0.23 & 8.34  & \textbf{0.39} & 4.15 \\
      mTACE & SOC\textsuperscript{a} & \textbf{0.18} & \textbf{6.85}  & 0.40 & \textbf{3.34} \\
      \bottomrule
    \end{tabular*}
  \end{minipage}\hfill
  \begin{minipage}[t]{0.49\linewidth}
    \centering
    \colorbox{green!10}{\strut\textbf{Noncollinear Fe-DeepSpin}}\\[3pt]
    \label{tab:noncollinear-fe-benchmark}
    \scriptsize
    \setlength{\tabcolsep}{1pt}
    \begin{tabular*}{\linewidth}{@{\extracolsep{\fill}}llrrrr}
      \toprule
      Model        & Type    & $E$       & $F$      & $S$   & $F^{\mathrm{mag}}$ \\
      \midrule
      DeePSPIN-DZP & SOC     & 8.60    & 111   & --    & 31.2 \\
      HotPP-spin   & non-SOC & 1.21    & 13.55 & --    & 6.94 \\
      mTACE        & non-SOC & 0.36    & 5.19  & 0.64  & \textbf{1.05} \\
      mTACE        & SOC     & \textbf{0.27}   & \textbf{5.10}  & \textbf{0.56}  & 1.06 \\
      \bottomrule
    \end{tabular*}
  \end{minipage}
  \vspace{3pt}
  \parbox{\linewidth}{\footnotesize\textsuperscript{a}For CrN, SOC models use e3nn without explicit time-reversal labels; independent global spin rotations and time reversal are imposed by data augmentation.}
\end{table}

\begin{figure}[hb]
  \centering
  \includegraphics[width=\linewidth]{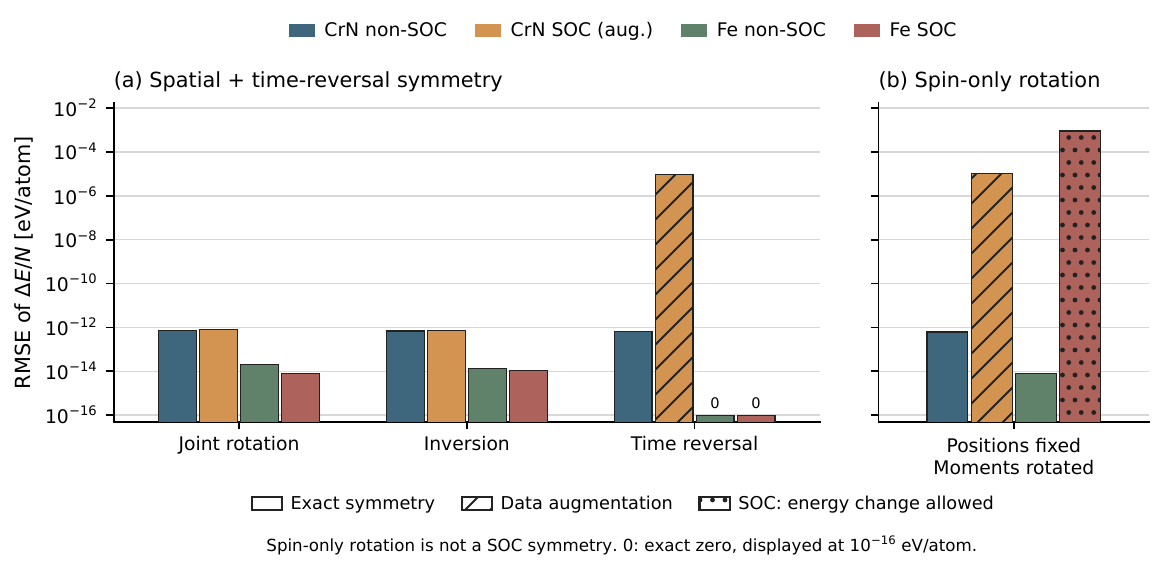}
  \caption{Energy symmetry tests in float64. (a) Joint spatial rotations, inversion, and time reversal. (b) Independent spin rotations, which need not preserve SOC energies. Colors identify models; diagonal hatching marks augmented symmetries and dotted hatching marks the SOC diagnostic. Exact zeros are displayed at a positive plotting floor on the logarithmic scale.}
  \label{fig:energy-symmetry}
\end{figure}

\section{Magnetic Benchmarks}
We evaluate mTACE on two public magnetic datasets: collinear CrN~\citep{CrN} and noncollinear Fe-DeepSpin~\citep{Fe-DeepSpin}. CrN and the Fe$_{16}$ subset of Fe-DeepSpin are each split into training, validation, and test sets at a ratio of $0.8:0.1:0.1$. For the CrN benchmark, we use e3nn without explicit time-reversal labels~\citep{e3nn}. When training the SOC architecture, we augment the data with spin rotations and time reversal. The Fe-DeepSpin labels were computed with SOC. However, the source study reports that SOC effects in elemental Fe are smaller than its intrinsic model errors and retained SOC primarily to validate the noncollinear spin-constrained DFT workflow. Following HotPP-spin~\citep{HotPP-spin}, we therefore train both SOC and non-SOC architectures on this dataset. The non-SOC model treats it as an exchange-dominated benchmark: it preserves relative-spin correlations while neglecting the unresolved SOC-induced coupling between spin orientation and the lattice. Table~\ref{tab:magnetic-benchmarks} compares mTACE with published mMTP, HotPP-spin, mMACE, and DeePSPIN-DZP results~\citep{CrN,HotPP-spin,mMACE,Fe-DeepSpin}. Across its variants, mTACE achieves the lowest RMSE for each evaluated property in this comparison. The reference results retain their original evaluation protocols; Appendix~\ref{app:training-hyperparameters} details the splits and our training settings. The local $O(2)$ construction retains the theoretical angular scaling of its $SO(2)$ counterpart~\citep{eSCN}, as analyzed in Appendix~\ref{app:complexity}.

\paragraph{Symmetry tests}
\label{sec:equivariance-error}

Alongside the prediction errors in Table~\ref{tab:magnetic-benchmarks}, we evaluate energy invariance for the four models on their respective test sets in float64 precision, reporting the RMSE of the energy change per atom in Figure~\ref{fig:energy-symmetry}. Our local $O(2)$ construction guarantees equivariance in exact arithmetic; symmetries enforced by construction exhibit only small finite-precision residuals in our numerical tests. In contrast, the CrN SOC model, which learns time reversal and independent spin rotations through data augmentation, retains substantially larger deviations under both transformations. Thus, data augmentation does not guarantee exact physical symmetry, motivating explicit representation and coupling constraints. Independent spin rotations test an additional non-SOC symmetry: the energy change of the Fe SOC model under these rotations is allowed by SOC and is not an equivariance violation.

\section{Conclusion}
We introduce two complementary approaches to efficient parity-complete equivariant computation. Generalized Wigner-$6j$ convolution exactly recouples two-stage, three-input equivariant tensor products, replacing the first edge-level intermediate with reusable node features. Our parity-complete local $O(2)$ framework resolves reflection and time-reversal actions through compatible representations and operators, guaranteeing global $O(3)\times\mathbb Z_2^{\mathcal T}$. At fixed channel multiplicities, its local linear-and-gating convolution retains the $\mathcal O(L^3)$ angular scaling of its $SO(2)$ counterpart. EquivariantX supplies a closed set of reusable linear, tensor product, gating, and frame operations across architectures, while mTACE built on EquivariantX demonstrates accurate magnetic learning with distinct SOC and non-SOC formulations. Together, these developments make complete spatial-parity and time-reversal constraints compatible with efficient local computations. Dedicated $O(2)$ operators, including optimized fused kernels, could further improve efficiency, while applications involving external fields and additional physical degrees of freedom offer broader uses for the framework.

\section*{AI Use Statement}
Large language models (LLMs) were used as assistive tools during this project. They assisted with code development and debugging, refinement of experimental designs, editing and proofreading of the manuscript, and verification of the implementation and mathematical derivations for potential errors. The research questions, ideas, research direction, experimental decisions, interpretation of the results, and conclusions were independently developed and determined by the authors. All LLM-generated content and suggestions were critically reviewed and verified by the authors, who take full responsibility for the final content of this work.

% \subsection*{Ethics statement}

% (This section is \textbf{recommended} and does not count toward the page limit.)

% If authors feel that their paper submission raises questions regarding the Code
% of Ethics, they are encouraged to include a paragraph of Ethics Statement (at
% the end of the main text before references) to address potential concerns where
% appropriate. Topics include, but are not limited to, studies that involve human
% subjects, practices to data set releases, potentially harmful insights,
% methodologies and applications, potential conflicts of interest and sponsorship,
% discrimination/bias/fairness concerns, privacy and security issues, legal
% compliance, and research integrity issues (e.g., IRB, documentation, research
% ethics). This statement should not be more than 1 page.

% \subsection*{Reproducibility statement}
%   The code for mTACE and EquivariantX is available at \url{https://github.com/xvzemin/tace}.

% \subsubsection*{Author Contributions}
% If you'd like to, you may include  a section for author contributions as is done
% in many journals. This is optional and at the discretion of the authors.

% \section*{Acknowledgment}
% This work was supported by the National Natural Science Foundation of China NSFC (22433004 and 22403064), the open research fund of the Key Laboratory of Precision and Intelligent Chemistry, and ShanghaiTech University. We are also grateful for the computing time provided by the HPC Platform of ShanghaiTech University.

\bibliography{references}
\bibliographystyle{iclr2027_conference}

\newpage
\appendix

\section{Notation}
\label{app:notation}

\begingroup
\footnotesize
\setlength{\tabcolsep}{0pt}
\renewcommand{\arraystretch}{1.05}
\noindent
\begin{tabular}{@{}p{0.27\linewidth}@{\hspace{0.02\linewidth}}p{0.71\linewidth}@{}}
$O(d),\ SO(d)$ & Orthogonal and special orthogonal groups, $d\in\{2,3\}$.\\
$\mathbb Z_2^{\mathcal T},\ \mathcal T,\ \tau$ & Time-reversal group, operation, and sign.\\
$G_{\mathrm{SOC}},\ G_{\mathrm{non\text{-}SOC}}$ & SOC and non-SOC symmetry groups.\\
$Q,\ Q_r,\ Q_s$ & Joint, spatial, and spin transformations.\\
$H,\ g(Q,\bm n)$ & Stabilizer of $\bm e_y$ and the induced stabilizer element.\\
$\ell,\ p,\ t$ & Degree, inversion parity, and time parity.\\
$p_v,\ t_v$ & Parities of the input vector.\\
$m,\ q$ & $O(2)$ order and reflection character.\\
$\rho_{\ell,p},\ \rho_{\rm in},\ \rho_{\rm out}$ & Irrep and input/output representations.\\
$R_{\bm n},\ R_x,\ R_y,\ \bm e_y,\ I_3$ & Alignment and axis rotations, $y$-axis unit vector, and identity.\\
$R_m(\theta),\ \theta,\ \phi$ & Order-$m$ rotation, rotation angle, and azimuth.\\
$S_0,\ F$ & Reference reflection and its two-dimensional representation.\\
$J$ & Change-of-basis matrix.\\
$B_{\rm in},\ B_{\rm out}$ & Input/output basis transformations.\\
$\sigma_{\rm in},\ \sigma_{\rm out}$ & Restricted representations in the $O(2)$ basis.\\
$D^{(\ell)},\ D_{ij}$ & Degree-$\ell$ Wigner-$D$ matrix and edge-alignment representation.\\
$\Phi,\ \mathcal F_{\bm n}$ & $H$-equivariant map and induced $O(3)$-equivariant map.\\
$i,\ j,\ Z_i,\ \mathcal N(i)$ & Target, source, species, and neighborhood.\\
$\bm r_i,\ \bm r_{ij},\ \bm n$ & Position, relative position, and unit reference direction.\\
$\bm m_i,\ \bm v$ & Magnetic moment and generic vector.\\
$h,\ h_0,\ h_{\pm m},\ h_m,\ s$ & $O(3)$ or $O(2)$ features, depending on the context. \\
$h_i^{(t)},\ h_{ij}^{(t)}$ & Node features and edge-aligned input at layer $t$.\\
$\nu$ & Product-basis correlation order.\\
$\mathcal Y^{(\ell)},\ \mathcal R^{(\ell)},\ Y^{(\ell)}$ & Spherical, regular solid, and edge harmonics.\\
$\mathcal M_i,\ \mathcal M_{ij}$ & Magnetic node and edge attributes.\\
$\ell_i,\ \ell_j,\ L,\ L_{\rm mag}$ & Input, output, and cutoff degrees; $L$ also denotes the complexity cutoff.\\
$\widetilde{\mathcal M}_{ij}^{(t)}$ & Weighted magnetic edge representation.\\
$W_{Z_i}^{(t)},\ W_{ij}^{(t)}$ & Element embedding and magnetic Linear weights.\\
$\bm W_{\text{conv},ij}^{(t)},\ \odot$ & Convolution weights and Hadamard product.\\
$E,\ \bm F_i,\ \bm F_i^{\mathrm{mag}}$ & Energy, atomic force, and magnetic force.\\
$F,\ F^{\mathrm{mag}},\ S$ & Atomic-force, magnetic-force, and stress RMSEs.\\
$A_{\mathcal T},\ A_S,\ A_P$ & Time-reversal, spin-rotation, and inversion augmentations.\\
$L_{\mathrm{max}}$ & Node-feature degree cutoff for the product-basis output.\\
$\ell_{\mathrm{max}}$ & Degree cutoff for edge features and interaction intermediates.\\
$m_{\mathrm{max}}$ & Local $O(2)$ order cutoff.\\
$k,\ u$ & Cartesian rank and solid-harmonic split degree.\\
$f,\ f_i,\ m_{\mathrm{cut},Z,f},\ \mathcal D_{Z,f}$ & Fidelity indices, magnetic cutoff, and training-set indices.\\
$u_i^2,\ \widetilde m_i$ & Scaled squared moment and transformed magnitude.\\
$T_n,\ T,\ N_{\rm mag}$ & Chebyshev polynomial, radial basis, and basis size.\\
$E_{\rm 1b},\ W_{Z,f,n}$ & One-body energy and element/fidelity coefficients.\\
$\pi,\ \ell_1,\ \ell_2,\ \ell_3$ & Coupling path and input degrees.\\
$a_j^{(\ell_3)},\ \ell_{12},\ \ell_{23}$ & Additional representation and intermediate degrees.\\
$[\,\cdot\,]_L$ & Projection onto degree $L$.\\
$\Phi_{ij,L}^{(\ell_{12})},\ g_{j,\pi}^{(\ell_{23})},\ A_{i,\pi}^{(L)}$ & Edge contraction, recoupled intermediate, and path aggregate.\\
$\Gamma,\ \widehat\Gamma$ & Wigner-$6j$ coefficient and normalized coefficient.\\
$s^{\rm dir},\ s^{\rm rec},\ \bar s^{\rm dir},\ \bar s^{\rm rec}$ & Path-normalization factors.\\
$W_{ij,\pi}^{\rm edge},\ W_{ij,\pi}^{\rm extra}$ & Reference tensor-product weights.\\
\end{tabular}
\endgroup

\clearpage

\section{Symmetry Foundations}
\label{sec:methods}

For the real-valued tensor features considered here, an irrep of $O(3)\times\mathbb Z_2^{\mathcal T}$ is labeled by an angular degree $\ell$, a spatial parity $p$, and a time-reversal parity $t$. We write this as $\ell pt$, representing the signs $p,t\in\{-1,+1\}$ by \texttt{o} and \texttt{e}, respectively. The first suffix always refers to spatial inversion and the second to time reversal. The full set of $O(3)\times\mathbb Z_2^{\mathcal T}$ irreps is therefore
\begin{equation}
  \bigoplus_{\ell=0}^{\infty}
  \left(
  \ell\texttt{ee}\oplus \ell\texttt{eo}\oplus
  \ell\texttt{oe}\oplus \ell\texttt{oo}
  \right).
  \label{eq:o3-time-irreps}
\end{equation}
Tensor-product paths obey the parity-selection rules $p_{\mathrm{out}}=p_1p_2$ and $t_{\mathrm{out}}=t_1t_2$, together with the angular-momentum rule $|\ell_1-\ell_2|\le \ell_{\mathrm{out}}\le \ell_1+\ell_2$.

\paragraph{Polar and axial vectors.}
Let $Q\in O(3)$ and let $\mathcal T$ denote time reversal.  A relative position is a time-even polar vector, whereas a noncollinear magnetic moment is a time-odd axial vector:
\begin{equation}
  \begin{aligned}
    Q:&\quad
    \bm r_{ij}\mapsto Q\bm r_{ij},
    &\bm m_i&\mapsto\det(Q)Q\bm m_i,\\
    \mathcal T:&\quad
    \bm r_{ij}\mapsto\bm r_{ij},
    &\bm m_i&\mapsto-\bm m_i.
  \end{aligned}
  \qquad
  \bm r_{ij}\sim\texttt{1oe},\quad
  \bm m_i\sim\texttt{1eo}.
  \label{eq:vector-irreps}
\end{equation}
After suppressing the time label, $\bm r_{ij}$ and $\bm m_i$ reduce to the familiar $O(3)$ irreps \texttt{1o} and \texttt{1e}, respectively.  In particular, spatial inversion reverses $\bm r_{ij}$ but leaves $\bm m_i$ unchanged, while time reversal has exactly the opposite action.

\paragraph{Spherical harmonics.}
Consider a nonzero vector $\bm v\sim 1p_vt_v$, identifying \texttt{e} with $+1$ and \texttt{o} with $-1$. The degree-$\ell$ spherical harmonic is the highest-degree irreducible component of $\widehat{\bm v}^{\otimes\ell}$ and hence transforms as $\mathcal Y^{(\ell)}(\widehat{\bm v})\sim\ell\,p_v^\ell t_v^\ell$. For relative positions and magnetic moments, respectively, this gives
% \begin{equation}
%   \begin{aligned}
%     \mathcal Y(\widehat{\bm r}_{ij})
%     &\sim
%     \texttt{0ee}\oplus\texttt{1oe}\oplus\texttt{2ee}
%     \oplus\texttt{3oe}\oplus\cdots,\\
%     \mathcal R({\bm m}_i)
%     &\sim
%     \texttt{0ee}\oplus\texttt{1eo}\oplus\texttt{2ee}
%     \oplus\texttt{3eo}\oplus\cdots.
%   \end{aligned}
%   \label{eq:position-magnetic-harmonics}
% \end{equation}
\begin{equation}
    \mathcal Y(\widehat{\bm r}_{ij})
    \sim
    \texttt{0ee}\oplus\texttt{1oe}\oplus\texttt{2ee}
    \oplus\texttt{3oe}\oplus\cdots,\quad
    \mathcal R({\bm m}_i)
    \sim
    \texttt{0ee}\oplus\texttt{1eo}\oplus\texttt{2ee}
    \oplus\texttt{3eo}\oplus\cdots.
  \label{eq:position-magnetic-harmonics}
\end{equation}
All degrees constructed from an axial vector are spatially even, while their time parity alternates with $\ell$. The corresponding regular solid harmonics $\mathcal R^{(\ell)}(\bm v)=\|\bm v\|^\ell \mathcal Y^{(\ell)}(\widehat{\bm v})$ carry the same spatial and temporal parities, while providing a smooth extension at $\bm v=\bm 0$.

\paragraph{With and without spin--orbit coupling.}
With spin--orbit coupling (SOC), the same rotation acts on the structure and magnetic moments. In the absence of an external time-reversal-breaking field, the energy transforms as the invariant scalar \texttt{0ee}. Equivalently, for every $Q\in O(3)$ and $\tau\in\{+1,-1\}\simeq\mathbb Z_2^{\mathcal T}$,
\begin{equation}
  E\!\left(
  \{Q\bm r_i,\tau\det(Q)Q\bm m_i,Z_i\}
  \right)
  =E\!\left(\{\bm r_i,\bm m_i,Z_i\}\right),
  \qquad
  G_{\mathrm{SOC}}=O(3)\times\mathbb Z_2^{\mathcal T}.
  \label{eq:energy-o3}
\end{equation}
Here, $\tau=-1$ denotes time reversal. Because the same $Q$ acts on both the spatial and spin degrees of freedom, this symmetry allows the energy to depend on the orientation of the magnetic moments relative to the lattice. Without SOC, spatial and spin rotations are independent. For $Q_r\in O(3)$ and $Q_s\in SO(3)$, the symmetry is instead
\begin{equation}
  \left.
  \begin{gathered}
    E(\{Q_r\bm r_i,Q_s\bm m_i,Z_i\})
    =E(\{\bm r_i,\bm m_i,Z_i\})
    \\
    E(\{\bm r_i,-\bm m_i,Z_i\})
    =E(\{\bm r_i,\bm m_i,Z_i\})
  \end{gathered}
  \right\},
  \qquad
  \begin{aligned}
    G_{\mathrm{non\text{-}SOC}}
    &=O(3)_{\mathrm{space}}\times SO(3)_{\mathrm{spin}}
    \times\mathbb Z_2^{\mathcal T}\\
    &\simeq O(3)_{\mathrm{space}}\times O(3)_{\mathrm{spin}}.
  \end{aligned}
  \label{eq:non-soc}
\end{equation}

\section{Generalized Wigner-\texorpdfstring{$6j$}{6j} Convolution}
\label{sec:wigner6j}

\subsection{Definition}

For notational brevity, parity and channel indices are omitted throughout this section. The \emph{generalized Wigner-6j convolution} couples three factors: a source node feature $h_j^{(\ell_2)}$, an edge spherical harmonic $Y^{(\ell_1)}(\widehat{\bm r}_{ij})$, and an independently supplied node representation $a_j^{(\ell_3)}$. The additional representation may contain regular solid harmonics of an applied electric-field direction or an atomic magnetic moment. More generally, it may be any invariant or equivariant node input.  An invariant input may enter the convolution weights, whereas a nonzero-degree input must enter the angular contraction if its orientation is to affect the message.  In the direct coupling order, the node feature and edge spherical harmonic first produce degree $\ell_{12}$, after which the result is coupled to the additional node representation:
\begin{equation}
  \Phi_{ij,L}^{(\ell_{12})}
  =
  \left[
    \left[
      h_j^{(\ell_2)}\otimes
      Y^{(\ell_1)}(\widehat{\bm r}_{ij})
    \right]_{\ell_{12}}
    \otimes
    a_j^{(\ell_3)}
  \right]_L.
  \label{eq:direct-tree}
\end{equation}
Here $[\,\cdot\,]_L$ denotes projection onto the degree-$L$ irreducible component.  Each possible coupling path is labeled by the following tuple:
\begin{equation}
  \pi=(\ell_1,\ell_2,\ell_{12},\ell_3,L).
  \label{eq:global-path}
\end{equation}
For repeated irreps, $\pi$ also identifies the input index, which we omit here for simplicity. The implementation uses channel-preserving \texttt{uvu} tensor products, with unit multiplicity for the edge harmonic and additional node representation. This order is costly because the first product is evaluated after source gather.  Its intermediate lives on edges, and the second product further expands edge computation.  To make the problem more concrete, we can interpret $a_j^{(\ell_3)}$ here as the regular solid harmonics of the noncollinear magnetic moment. Such a convolution scheme is in fact an essential ingredient in the convolution design of mMACE~\citep{mMACE}. The generalized Wigner-$6j$ convolution rearranges the coupling order so that terms involving only the source atom $j$ are evaluated before source gather. This removes the first edge-level intermediate. Avoiding storage of the final edge message additionally requires a fused tensor-product and aggregation backend, such as OpenEquivariance~\citep{oeq} and CuEquivariance~\citep{cueq}.

\subsection{Exact Wigner-\texorpdfstring{$6j$}{6j} recoupling}

We first construct the source-node intermediate
\begin{equation}
  g_{j,\pi}^{(\ell_{23})}
  =
  \left[
    h_j^{(\ell_2)}
    \otimes
    a_j^{(\ell_3)}
  \right]_{\ell_{23}},
  \label{eq:node-extra-intermediate}
\end{equation}
and then couple it to the edge harmonic:
\begin{equation}
  \Phi_{ij,L}^{(\ell_{12})}
  =
  \sum_{\ell_{23}}
  \Gamma^{L}_{\ell_1\ell_2\ell_{12}\ell_3\ell_{23}}
  \left[
    g_{j,\pi}^{(\ell_{23})}
    \otimes Y^{(\ell_1)}(\widehat{\bm r}_{ij})
  \right]_L.
  \label{eq:recoupled-tree}
\end{equation}
For this operand order and the component-normalized CG convention used by \texttt{e3nn}~\citep{e3nn}, the recoupling coefficient is
\begin{equation}
  \Gamma^{L}_{\ell_1\ell_2\ell_{12}\ell_3\ell_{23}}
  =
  (-1)^{\ell_1+\ell_3+\ell_{12}+\ell_{23}}
  \sqrt{(2\ell_{12}+1)(2\ell_{23}+1)}
  \begin{Bmatrix}
    \ell_1 & \ell_2 & \ell_{12}\\
    \ell_3 & L & \ell_{23}
  \end{Bmatrix}.
  \label{eq:wigner6j}
\end{equation}
Here the braces denote the Wigner-$6j$ symbol, evaluated using \texttt{SymPy}~\citep{sympy}. For each original path $\pi$, Eqs.~\eqref{eq:direct-tree} and \eqref{eq:recoupled-tree} are exactly equivalent when all allowed $\ell_{23}$ with nonzero coefficients are retained. Matching the implemented tensor products also requires accounting for path normalization. Our reference uses \texttt{irrep\_normalization=component} and \texttt{path\_normalization=element}. Element normalization depends on the contributing paths and their multiplicities, so its factors can differ between the two coupling trees. Let $s_{\pi}^{\rm dir}$ and $s_{\pi,\ell_{23}}^{\rm rec}$ denote the products of the two generated instruction-normalization factors along the corresponding paths. These are fixed factors, not learned weights. Removing the component-normalization factors gives the element scales
\begin{equation}
  \bar{s}_{\pi}^{\rm dir}
  =\frac{s_{\pi}^{\rm dir}}
  {\sqrt{(2\ell_{12}+1)(2L+1)}},
  \qquad
  \bar{s}_{\pi,\ell_{23}}^{\rm rec}
  =\frac{s_{\pi,\ell_{23}}^{\rm rec}}
  {\sqrt{(2\ell_{23}+1)(2L+1)}}.
  \label{eq:element-scales}
\end{equation}
The coefficient used by the recoupled implementation is consequently
\begin{equation}
  \widehat{\Gamma}^{L}_{\ell_1\ell_2\ell_{12}\ell_3\ell_{23}}
  =
  \Gamma^{L}_{\ell_1\ell_2\ell_{12}\ell_3\ell_{23}}
  \frac{\bar{s}_{\pi}^{\rm dir}}
  {\bar{s}_{\pi,\ell_{23}}^{\rm rec}}.
  \label{eq:normalized-wigner6j}
\end{equation}
Let $W_{ij,\pi}^{\rm edge}$ and $W_{ij,\pi}^{\rm extra}$ be the channel-wise scalar weights for the first and second reference tensor products, respectively. In the magnetic interaction, both are generated on edges from the concatenation of the spatial edge features and the source magnetic radial basis $T(\widetilde m_j)$. Since scalar weights commute with the angular contractions, their product is applied in the final node-edge tensor product. The node-node tensor product uses fixed unit weights. Using the implementation-normalized tensor products for $g$ and the final contraction, the aggregated output of each original path is
\begin{equation}
  A_{i,\pi}^{(L)}
  =
  \sum_{j\in\mathcal{N}(i)}
  \sum_{\ell_{23}}
  W_{ij,\pi}^{\rm edge}W_{ij,\pi}^{\rm extra}
  \widehat{\Gamma}^{L}_{\ell_1\ell_2\ell_{12}\ell_3\ell_{23}}
  \left[
    g_{j,\pi}^{(\ell_{23})}
    \otimes Y^{(\ell_1)}(\widehat{\bm r}_{ij})
  \right]_L.
  \label{eq:wigner6j-conv}
\end{equation}
Distinct original paths retain separate weights and output channels; a subsequent equivariant linear map mixes these outputs.

\subsection{Intermediate reuse and execution}

The implementation evaluates each distinct combination of source-feature copy, additional-attribute copy, and intermediate irrep once per node. Original paths that share this combination reuse the same $g_{j,\pi}^{(\ell_{23})}$. The intermediates are stored in irrep-sorted order, making equal irrep types contiguous without summing independent channels or requiring a runtime permutation. Recoupling preserves every original path and includes every allowed intermediate contribution with a nonzero coefficient. It removes one edge-level tensor product stage and reuses node intermediates, but does not eliminate the coupling-path growth of global CGTPs. This motivates the local $O(2)$ formulation developed in the main text.

\section{Data Augmentation}
\label{sec:symmetry-catalogue}

With SOC, spatial and spin transformations are locked: a single $Q\in O(3)_{\mathrm{space+spin}}$ acts on positions as $\bm r_i\mapsto Q\bm r_i$ and on axial magnetic moments as $\bm m_i\mapsto\det(Q)Q\bm m_i$.  Without SOC, the two sectors transform independently.  An arbitrary $Q_r\in O(3)_{\mathrm{space}}$ acts on the positions, while an independent $Q_s\in SO(3)_{\mathrm{spin}}$ acts on all magnetic moments.  Their actions are $\bm r_i\mapsto Q_r\bm r_i$ and $\bm m_i\mapsto Q_s\bm m_i$. The SOC architecture admits the broader function space: it may depend on the orientation of the magnetic moments relative to the lattice.  This is required for SOC data but is an unwanted degree of freedom for a non-SOC target. Thus, when the more general SOC architecture is applied to non-SOC data, data augmentation is used to encourage the appropriate non-SOC symmetry when it is not enforced by the architecture. Let $A_{\mathcal T}$ denote time reversal, $A_P$ spatial inversion, and $A_S$ a spin rotation. Their actions on positions, magnetic moments, atomic forces, and magnetic forces are
\begin{align}
  A_{\mathcal T}:&\quad
  (\bm r_i,\bm m_i,\bm F_i,\bm F_i^{\mathrm{mag}})
  \mapsto(\bm r_i,-\bm m_i,\bm F_i,-\bm F_i^{\mathrm{mag}}),
  \\
  A_P:&\quad
  (\bm r_i,\bm m_i,\bm F_i,\bm F_i^{\mathrm{mag}})
  \mapsto(-\bm r_i,\bm m_i,-\bm F_i,\bm F_i^{\mathrm{mag}}),
  \\
  A_S(R):&\quad
  (\bm r_i,\bm m_i,\bm F_i,\bm F_i^{\mathrm{mag}})
  \mapsto(\bm r_i,R\bm m_i,\bm F_i,R\bm F_i^{\mathrm{mag}}),
  \qquad R\in SO(3).
\end{align}
For non-SOC data without an external time-reversal-breaking field, the energy is unchanged under all three transformations. For SOC data, the preferred choice is the SOC architecture with complete spatial parity and explicit time-reversal parity; its group is $O(3)_{\mathrm{space+spin}}\times\mathbb Z_2^{\mathcal T}$ and no augmentation is needed.  For non-SOC data, the preferred choice is the explicit non-SOC architecture, whose group is $O(3)_{\mathrm{space}}\times SO(3)_{\mathrm{spin}} \times\mathbb Z_2^{\mathcal T}$ and likewise requires no augmentation.  An SOC architecture should be used for non-SOC data only when its broader function space is deliberately retained and constrained through augmentation.

\section{Magnetic Cutoff and Basis Normalization}
\label{sec:magnetic-normalization}

\paragraph{Magnetic cutoff.}
Let $m_{\mathrm{cut},Z,f}>0$ denote the magnetic cutoff for element $Z$ and fidelity $f$. For each fidelity, it may be supplied as a positive scalar shared by all elements or as an element-dependent mapping; supplied values are used directly, without further rescaling. Otherwise, the cutoff is estimated independently from that fidelity's training statistics~\citep{mMACE}:
\begin{equation}
  m_{\mathrm{cut},Z,f}=1.2\max_{i\in\mathcal D_{Z,f}}\|\bm m_i\|+0.1\,\mu_{\mathrm B},
  \label{eq:magnetic-cutoff}
\end{equation}
where $\mathcal D_{Z,f}$ indexes the training-set moments of element $Z$ at fidelity $f$, and magnetic moments and cutoffs are expressed in Bohr magnetons $\mu_{\mathrm B}$. A zero or missing element entry in an available statistics mapping gives a cutoff of $0.1\,\mu_{\mathrm B}$. Cutoffs are stored separately for each element and fidelity and remain fixed during training.

\paragraph{Transformed magnitude features.}
The magnetic cutoff controls only the radial magnitude map, not the angular basis. Let $f_i$ denote the fidelity of atom $i$'s configuration. The implementation forms the squared scaled magnitude $u_i^2=\sum_{\alpha=x,y,z}(m_{i,\alpha}/m_{\mathrm{cut},Z_i,f_i})^2$. The transformed scalar magnitude $\widetilde m_i$ and magnetic node features $T(\widetilde m_i)$ are
\begin{equation}
  \widetilde m_i=1-2\min(u_i^2,1)\in[-1,1],
  \qquad
  T(\widetilde m_i)=\bigl(T_1(\widetilde m_i),\ldots,T_{N_{\rm mag}}(\widetilde m_i)\bigr),
  \label{eq:magnetic-radial-basis}
\end{equation}
where $T_n$ is the first-kind Chebyshev polynomial, evaluated by the recurrence $T_0(x)=1$, $T_1(x)=x$, and $T_{n+1}(x)=2xT_n(x)-T_{n-1}(x)$ for $n\geq1$. The parameter $N_{\rm mag}$ counts the returned nonconstant basis functions: $T_1,\ldots,T_{N_{\rm mag}}$. The radial constant $T_0$ is excluded from the interactions.

\paragraph{Smoothness and saturation.}
Saturation at $\|\bm m_i\|=m_{\mathrm{cut},Z_i,f_i}$ is intentional: it keeps the Chebyshev input within $[-1,1]$ beyond the chosen magnetic range. Inside the open domain $\|\bm m_i\|<m_{\mathrm{cut},Z_i,f_i}$, the magnitude map is a polynomial in the magnetic components, the Chebyshev basis is polynomial, and the regular solid harmonics are Cartesian polynomials. With smooth network activations, the energy is therefore smooth with respect to the magnetic moments, including at zero. Magnetic-structure optimization confined to this domain encounters no discontinuity introduced by these basis functions. The hard saturation is continuous but generally not differentiable at the cutoff; crossing it can produce a jump in magnetic forces. The released magnetic radial implementations of mMACE~\citep{mMACE} and HotPP-spin~\citep{HotPP-spin} likewise hard-clamp the moment magnitude at an upper cutoff. Such possible force discontinuities when crossing the cutoff are therefore an expected consequence of hard saturation, rather than a limitation specific to the local $O(2)$ formulation. The cutoff should therefore be chosen above the moment range explored in optimization.

\paragraph{Angular basis.}
The angular representation is evaluated on the original magnetic vector $\bm m_i$, without applying the magnetic cutoff:
\begin{equation}
  \mathcal M_i^{(\ell)}=\mathcal R^{(\ell)}(\bm m_i)
  =\|\bm m_i\|^\ell\mathcal Y^{(\ell)}(\widehat{\bm m}_i),
  \qquad 0\leq \ell\leq L_{\rm mag}.
  \label{eq:magnetic-solid-harmonics}
\end{equation}
The implementation uses integral normalization, multiplying the component-normalized solid harmonics by $1/\sqrt{4\pi}$. It evaluates the Cartesian polynomials directly, without division by the moment magnitude or element- or fidelity-dependent angular rescaling. Each $\ell>0$ irrep is therefore homogeneous of degree $\ell$ and vanishes smoothly at zero, while the angular scalar $\mathcal M_i^{(0)}=1/\sqrt{4\pi}$ is retained. The $\ell>0$ angular features are unbounded and scale as $\|\bm m_i\|^\ell$.

\paragraph{One-body magnetic readout.}
To make the physical interpretation of the energy decomposition explicit, one may additionally impose the readout constraints proposed in mMACE~\citep{mMACE}; we do not discuss this extension further here. When enabled, the magnitude-only one-body readout is an element-dependent linear map with one output head per fidelity, no hidden layers, and no separate bias. Selecting the head for each configuration gives
\begin{equation}
  E_{\rm 1b}=\sum_i\left(
    W_{Z_i,f_i,0}+\sum_{n=1}^{N_{\rm mag}}W_{Z_i,f_i,n}T_n(\widetilde m_i)
  \right).
  \label{eq:magnetic-one-body}
\end{equation}
The constant input $T_0=1$ is appended only at readout time and supplies the independent coefficient $W_{Z_i,f_i,0}$. The element- and fidelity-dependent atomic-energy baseline is added separately.

\section{Computational Complexity}
\label{app:complexity}

EquivariantX supports arbitrary direct sums of \texttt{e3nn} irreps with entry-dependent multiplicities through degree-wise Wigner-$D$ rotations. Entries with equal multiplicity and distinct degrees are grouped into shared rotation applications to reduce kernel-launch overhead. At a fixed degree, the proper-rotation matrices are independent of spatial and time-reversal parities and can be reused across these sectors. The fixed basis change can also be absorbed into the corresponding frame matrices, together with its inverse in the local-to-global transformation, rather than applied separately. The additional global CGTP coupling paths introduced by complete $O(3)$ and time-reversal labels further motivate the local-frame approach, which retains order-wise operations and cubic theoretical angular complexity instead of explicitly evaluating all global angular couplings.

For maximum angular degree $L$ and fixed channel multiplicity, a dense CGTP has angular complexity $\mathcal O(L^6)$; exploiting the sparsity of the Clebsch--Gordan coefficients reduces it to $\mathcal O(L^5)$. In contrast, the $SO(2)$ formulation of eSCN~\citep{eSCN} has $\mathcal O(L^3)$ angular complexity. The analysis for our local $O(2)$ frame proceeds analogously. 

With all local orders retained, the degree-$\ell$ Wigner-$D$ matrix acts on $2\ell+1$ angular features, giving quadratic cost in $2\ell+1$ per rotation. At local order $m$, each degree $\ell=m,\ldots,L$ contributes one copy of the irrep, so a Linear mixes channel widths proportional to $L-m+1$. Summing over degrees and orders gives Wigner-$D$ rotations: $\mathcal O\!\left(\sum_{\ell=0}^{L}(2\ell+1)^2\right)=\mathcal O(L^3)$, and Local Linear: $\mathcal O\!\left(\sum_{m=0}^{L}(L-m+1)^2\right)=\mathcal O(L^3)$. The one- and two-dimensional real irrep dimensions contribute only constant factors. Element-wise activations and gate multiplication cost $\mathcal O(L^2)$; generating the scalar gates from local invariant features costs at most $\mathcal O(L^3)$. For unnatural-parity representations, the additional fixed orthogonal basis change exchanges and changes signs of the two real features at each positive order. Its cost is $\mathcal O(\sum_{\ell=0}^{L}\ell)=\mathcal O(L^2)$, leaving the total theoretical angular complexity $\mathcal O(L^3)$. Complete spatial parity and time-reversal labels change block multiplicities without changing this exponent. 

\section{EquivariantX}
\label{sec:eqx-example}

\subsection{Local O(2) Convolution}

Listing~\ref{lst:eqx-local-frame} constructs a minimal edge-aligned local $O(2)$ convolution and checks its global $O(3)$ equivariance under rotations, inversion, and their composition. The global input/output irreps and test transformations are created with \texttt{e3nn.o3}; \texttt{eqx.o2} provides the Wigner-$D$ matrices, local frames, and \texttt{linear\_up} $\rightarrow$ \texttt{o2\_gate} $\rightarrow$ \texttt{linear\_down} pipeline. The example includes both spatial parities at every degree up to $\ell=2$, with $64$ channels per global irrep, so it exercises the basis changes for unnatural-parity features and their mixing with natural-parity features. Edge-gathered node features and messages have shape \texttt{(num\_edges, 18, 64)} and are flattened to \texttt{(num\_edges, 1152)} in \texttt{ir\_mul} order. The restricted representation has $192$ copies each of \texttt{0e} and \texttt{0o}, $256$ copies of \texttt{1m}, and $128$ copies of \texttt{2m}. The gate applies SiLU to \texttt{0e}, tanh to \texttt{0o}, and sigmoid-activated \texttt{0e} gates to positive-order irreps. Rotations act on the angular axis, and all computations use float64.

\begin{lstlisting}[
  style=eqxpython,
  caption={Executable global-$O(3)$/local-$O(2)$ equivariance check.},
  label={lst:eqx-local-frame}
]
import torch
from e3nn import o3
from eqx import o2

torch.set_default_dtype(torch.float64)
torch.manual_seed(0)
num_edges = 16
num_channels = 64
lmax = 2
angular_irreps = o3.Irreps("0e + 0o + 1o + 1e + 2e + 2o")
irreps_in = o3.Irreps([(num_channels, ir) for _, ir in angular_irreps])
irreps_out = irreps_in

local_frame_in = o2.LocalFrame(irreps_in, mmax=lmax)
local_frame_out = o2.LocalFrame(irreps_out, mmax=lmax, reverse=True)
wignerd = o2.WignerD(mmax=lmax, lmax=lmax)

o2_gate = o2.Gate(
    f"{3 * num_channels}x0e + {3 * num_channels}x0o",
    [torch.nn.SiLU(), torch.nn.Tanh()],
    f"{6 * num_channels}x0e", [torch.nn.Sigmoid()],
    f"{4 * num_channels}x1m + {2 * num_channels}x2m",
)
linear_up = o2.Linear(local_frame_in.irreps_out, o2_gate.irreps_in)
linear_down = o2.Linear(o2_gate.irreps_out, local_frame_out.irreps_out)


@torch.no_grad()
def o2_interaction(node_feats, edge_vector):
    wigner, wigner_inv = wignerd(edge_vector)
    node_features = local_frame_in.to_local(node_feats.flatten(-2), wigner)
    projected = linear_up(node_features)
    hidden = o2_gate(projected)
    message = linear_down(hidden)
    return local_frame_out.to_global(message, wigner_inv).view_as(node_feats)


edge_vector = torch.randn(num_edges, 3)
node_feats = torch.randn(num_edges, angular_irreps.dim, num_channels)
rotation = o3.rand_matrix()

message = o2_interaction(node_feats, edge_vector)
for name, Q in (
    ("rotation", rotation),
    ("inversion", -torch.eye(3)),
    ("rotation + inversion", -rotation),
):
    representation = angular_irreps.D_from_matrix(Q)
    transformed_message = o2_interaction(
        representation @ node_feats,
        edge_vector @ Q.T,
    )
    expected_message = representation @ message
    torch.testing.assert_close(
        transformed_message, expected_message,
        atol=1.0e-10, rtol=1.0e-10,
    )
    error = (transformed_message - expected_message).abs().max().item()
    print(f"{name}: max error = {error:.3e}")
\end{lstlisting}

\subsection{O(2) Tensor product}

Listing~\ref{lst:eqx-tensor-product} uses \texttt{o2.TensorProduct} directly, without frame rotations, to evaluate $\texttt{1m}\otimes\texttt{1m}=\texttt{0e}\oplus\texttt{0o}\oplus\texttt{2m}$ with $64$ channels per irrep. Each instruction specifies \texttt{(i\_in1, i\_in2, i\_out, mode, train)}: the indices select irrep entries, \texttt{uuu} couples matching channels, and \texttt{True} enables a learnable weight for each channel on that path. By default, these weights are stored internally and shared across the batch; the Clebsch--Gordan coefficients are fixed. Using \texttt{uvw} instead enables dense mixing of input and output channels. Each irrep entry uses flattened \texttt{ir\_mul} storage, giving input shapes \texttt{(16, 128)} and output shape \texttt{(16, 256)}.

\begin{lstlisting}[
  style=eqxpython,
  caption={Channel-wise $O(2)$ tensor product with learnable path weights.},
  label={lst:eqx-tensor-product}
]
import torch
from eqx import o2

tp = o2.TensorProduct(
    irreps_in1="64x1m",
    irreps_in2="64x1m",
    irreps_out="64x0e + 64x0o + 64x2m",
    instructions=[
        (0, 0, 0, "uuu", True),  # 1m x 1m -> 0e
        (0, 0, 1, "uuu", True),  # 1m x 1m -> 0o
        (0, 0, 2, "uuu", True),  # 1m x 1m -> 2m
    ],
)
h1 = torch.randn(16, 2, 64).flatten(-2)
h2 = torch.randn(16, 2, 64).flatten(-2)
h_out = tp(h1, h2)
assert h_out.shape == (16, 256)
print(h_out.shape)
\end{lstlisting}

\section{Training Hyperparameters}
\label{app:training-hyperparameters}

Table~\ref{tab:training-hyperparameters} summarizes the $O(2)$-mTACE configurations used for CrN and Fe-DeepSpin. We use training, validation, and test sets with an $80\%/10\%/10\%$ split; the validation set is reserved for model selection and the test set for final evaluation. For Fe-DeepSpin, this split applies only to Fe$_{16}$. HotPP-spin~\citep{HotPP-spin} uses 30,806 Fe$_{16}$ configurations for training and 3,422 for testing, approximately a $90\%/10\%$ train/test split. The mMACE CrN results~\citep{mMACE} are evaluated on the original benchmark test set~\citep{CrN}; its paper does not specify a numerical train/validation/test split for this benchmark. We regard our strict three-way split as a conservative and fair literature comparison. This does not establish identical test samples across studies, and the reported baseline errors retain their original evaluation protocols. All our models were trained in float32 precision on a single NVIDIA GeForce RTX 4090 GPU.
\begin{table}[h!]
  \caption{Hyperparameters used for our training runs on the two magnetic benchmarks.}
  \label{tab:training-hyperparameters}
  \centering
  % \small
  \setlength{\tabcolsep}{4pt}
  \begin{tabular}{lcc}
    \toprule[1.2pt]
    Hyperparameter                   & CrN               & Fe-DeepSpin \\
    Training split                   & $80\%$            & $80\%$ \\
    Validation split                 & $10\%$            & $10\%$ \\
    Test split                       & $10\%$            & $10\%$ \\
    Optimizer                        & AdamW             & AdamW \\
    Initial learning rate            & $1\times10^{-3}$  & $1\times10^{-3}$ \\
    Learning-rate scheduler          & ReduceLROnPlateau & ReduceLROnPlateau \\
    Learning-rate reduction factor   & $0.5$             & $0.5$ \\
    Learning-rate patience           & $25$              & $10$ \\
    Early-stopping patience          & $50$              & $20$ \\
    Total batch size                 & $4$               & $8$ \\
    Weight decay                     & $1\times10^{-8}$  & $1\times10^{-8}$ \\
    Loss properties                  & $E/F/S/F^{\mathrm{mag}}$ & $E/F/S/F^{\mathrm{mag}}$ \\
    Loss coefficients                & $1/5/1/10$        & $1/5/1/10$ \\
    Loss types                       & MSE/MSE/MSE/MSE   & MSE/MSE/MSE/MSE \\
    Gradient clipping norm threshold & $10$              & $10$ \\
    Model EMA decay                  & $0.999$           & $0.999$ \\
    \midrule[0.6pt]
    Spatial cutoff radius (\AA)      & $5$               & $5$ \\
    Number of layers                 & $6$               & $4$ \\
    Interaction channels             & $64$              & $48$ \\
    Radial Rotary Complex Attention  & Enabled           & Enabled \\
    $L_{\mathrm{max}}$               & $2$               & $2$ \\
    $\ell_{\mathrm{max}}$            & $2$               & $2$ \\
    $m_{\mathrm{max}}$               & $2$               & $2$ \\
    $L_{\rm mag}$                   & $2$               & $2$ \\
    Correlation order                & $2$               & $2$ \\
    \bottomrule[1.2pt]
  \end{tabular}
\end{table}

\end{document}